\documentstyle[aps,prc,floats,graphicx,amsmath,amssymb]{revtex}

\newcommand{\UCN}{_{\scriptscriptstyle UCN}}

\begin{document}

\draft
    
\title{Magnetic trapping of ultracold neutrons}

\author{C. R. Brome\thanks{Present address: Sloan Center for
Theoretical Neurobiology, University of California, San Francisco, San
Francisco, CA 94143}, J. S. Butterworth\thanks{Present address:
Institut Laue-Langevin, BP 156-6 rue Jules Horowitz, 38042, Grenoble
(Cedex 9), France}, S. N. Dzhosyuk, C. E. H. Mattoni, D. N. McKinsey,
and J. M. Doyle} \address{Harvard University, 17 Oxford Street,
Cambridge, MA 02138, USA}

\author{P. R. Huffman\thanks{Correspondence and requests for materials
should be addressed to P.R.H. (email: paul.huffman@nist.gov).}, M. S.
Dewey, and F. E. Wietfeldt} \address{National Institute of Standards
and Technology, 100 Bureau Drive, MS 8461, Gaithersburg, MD 20899,
USA}

\author{R. Golub and K. Habicht}
\address{Hahn-Meitner Institut, Glienicker Strasse 100, D-14109 Berlin, Germany} 

\author{G. L. Greene and S. K. Lamoreaux}
\address{University of California, Los Alamos National Laboratory, PO 
Box 1663, Los Alamos, NM 87545, USA}

\author{K. J. Coakley}
\address{National Institute of Standards and Technology, 325 
Broadway, MS 898.02, Boulder, CO 80305, USA}

\maketitle

\date{\today}

\begin{abstract}
Three-dimensional magnetic confinement of neutrons is reported. 
Neutrons are loaded into an Ioffe-type superconducting magnetic trap
through inelastic scattering of cold neutrons with $^4$He.  Scattered
neutrons with sufficiently low energy and in the appropriate spin
state are confined by the magnetic field until they decay.  The electron 
resulting from neutron decay produces scintillations in the liquid helium bath
that results in a pulse of extreme ultraviolet light.  This light is
frequency downconverted to the visible and detected.  Results are
presented in which $500 \pm 155$ neutrons are magnetically trapped in
each loading cycle, consistent with theoretical predictions.  The
lifetime of the observed signal, $660~\mathrm{s}~^{+290}_{-170}$~s, is
consistent with the neutron beta-decay lifetime.
\end{abstract}


\twocolumn
\narrowtext

\section{Introduction}
\label{sect:intro}

The beta-decay lifetime of the neutron is of interest in a number of
contexts.  It has a direct impact on cosmological modeling as an
input to the calculation of the production of light elements during
the Big Bang.  The isotopic ratios measured in extragalactic gas
clouds (in which isotopic abundances are thought to have remained
constant since the Big Bang) can be compared to Big Bang
Nucleosynthesis calculations in order to place limits on the ratio of
baryons to photons in the early universe.  In the calculation of the
expected $^{4}$He/$^{1}$H ratio, the dominant uncertainty is the
lifetime of the neutron\cite{Tur99,PDG}.

Beta decay of the neutron is both the simplest nuclear beta decay, and
more generally, the simplest of the charged-current weak interactions
in baryons.  Measurements of the weak interaction parameters can be
obtained using neutron beta decay with fewer and simpler theoretical
corrections than measurements using the beta decay of nuclei.  The
neutron beta decay rate is proportional to the quantity
$g_{v}^{2}+3g_{a}^{2}$ where $g_{v}$ and $g_{a}$ are the semileptonic
vector and axial-vector coupling constants.  To extract the coupling
constants uniquely from neutron beta-decay measurements requires
either an independent measurement of $g_{v}$ or $g_{a}$ or of the
(more experimentally accessible) ratio $g_{a}/g_{v} \equiv \lambda$.

The technique of magnetic trapping offers the promise of significant
improvements over previous measurements in the accuracy of the neutron
lifetime ($\tau_{n}$).  Past measurements have relied on two
techniques for measuring $\tau_{n}$: beam and closed-sample type
measurements.  At present, both techniques appear to be systematically
limited at the $10^{-3}$ level.

In a beam measurement, a well collimated cold neutron beam passes
continuously through a decay region of known volume.  Each neutron
decay occurring in that volume is counted.  At the end of the decay
region, the flux of the neutron beam is measured with a 1/$v$
weighting, where $v$ is the velocity.  The neutron lifetime is ideally
given by the ratio of the number of decays per unit time to the number
of neutrons in the decay region.  In the most precise measurement to
date using this method ($\tau_{n}=889.2~\mathrm{s} \pm 4.8$~s
\cite{Byr90,Byr96}), neutron decays were counted by detecting the
decay protons and the beam flux was measured by detecting
$\alpha$-particles emitted in the $n + \mathrm{^{10}B} \rightarrow
\mathrm{^{7}Li} + \alpha$ reaction.  Of the 4.8~s uncertainty, 3.4~s
is attributed to systematic effects in the measurement of the neutron
flux.  An improved version of this experiment (presently running at
the National Institute of Standards and Technology, NIST) aims for an
accuracy of less than two seconds\cite{nico}.

In the closed-sample measurements, neutrons with energies lower than
the interaction potential of the material (i.e.\ ultracold neutrons or
UCN) are loaded into a physical box and stored (via total reflection
of the neutrons from the material surfaces) for a variable length of
time before being counted.  The neutron lifetime can be extracted from
the dependence of the detected neutron population on the storage time. 
Several storage techniques have been used to make measurements of the
neutron lifetime.  The most precise storage measurements yielded
values of $\tau_{n} = 882.6~\mathrm{s} \pm 2.7$~s\cite{Mam93},
$888.4~\mathrm{s} \pm 3.3$~s\cite{Nes92} and $885.4~\mathrm{s} \pm
0.9~s \pm 0.4$~s\cite{Arz00}.

Experiments using this storage technique to measure the neutron
lifetime measure the total rate at which neutrons are lost from the
storage vessel, both by beta decay and by any other loss mechanisms. 
The accuracy of the lifetime measurement is limited either by unknown
loss mechanisms or by the uncertainty in corrections for large loss
mechanisms which arise from the interactions between the stored
neutrons and the walls of the storage vessel.  Most experiments use
storage data to extrapolate to the ideal case of no wall losses.  In
Ref.~\cite{Mam93}, the physical size of the storage vessel is varied,
and the measurements are extrapolated to an infinite volume.  In
Ref.~\cite{Nes92}, the storage time is measured as a function of UCN
velocity, and an extrapolation is made to zero velocity, or an
infinite time between collisions with the walls.  The most recent
measurement, \cite{Arz00}, uses two different storage-vessel sizes
with neutron detectors to measure the relative rate at which neutrons
are scattered out of the storage vessel in the two configurations.

A second storage technique uses the interaction of the magnetic moment
of the neutron with magnetic field gradients.  This technique was
first discussed by Vladimirski$\mathrm{\breve{i}}$ in
1961\cite{Vla61}, just two years after Zeldovich first discussed
material storage\cite{Zel59}.  Since that time there have been
several experiments to store neutrons magnetically with varying
degrees of success\cite{Abo86,Pau89,Nie83,Nes85}.

In the NESTOR (NEutron STOrage Ring) experiment, neutrons in a range
of velocities from 6~m~s$^{-1}$ to 20~m~s$^{-1}$ were stored in a
sextupole magnetic storage ring analogous to the storage rings used at
particle accelerators\cite{Pau89,Nes85}.  Magnetic field gradients
reflect neutrons at a glancing incidence, storing neutrons in certain
trajectories.  The storage ring technique relies on the stored
neutrons to remain in these trajectories.  Neutron loss from these
trajectories (attributed to betatron oscillation) limited this
measurement.

Three-dimensional magnetic trapping was later attempted with a
spherical hexapole trap\cite{Nie83}.  The lack of neutrons trapped in
this experiment was attributed to upscattering of neutrons from the
trap by phonons in their 1.2~K helium bath.  Another experiment used a
large ($3 \times 10^{5}$~cm$^{3}$) box with several current loops
providing the magnetic field for confinement along the floor and walls
(the top was ``closed'' by gravity).  Despite its size it was never
able to confine more than a few (of order one) neutrons at a time due
to its a trap depth of $150~\mu$K\cite{Abo86,Abo83}.

The use of three-dimensional magnetic confinement to measure the
neutron lifetime provides an environment that is free of the
systematic effects that have limited previous lifetime measurements. 
The magnetic trapping technique is not sensitive to variations in the
neutron flux; each decay is recorded as a function of time and both
the lifetime and initial number of neutrons can be extracted. 
Magnetic trapping also eliminates wall interactions and betatron
losses; neutrons are not confined by physical walls and are
energetically forbidden from leaving the trapping region.  These
advantages should allow for a significant improvement in the accuracy
of the neutron lifetime.  Given a sufficient neutron flux it should be
possible (using estimates of known loss mechanisms discussed below) to
measure $\tau_{n}$ at the $10^{-5}$ level.  The method of measuring
the neutron lifetime using magnetic trapping that is discussed in this
paper was proposed in 1994\cite{Doy94}.

In this work ultracold neutrons are produced, magnetically trapped,
and detected as they decay.  The experimental concepts are discussed
in section~\ref{sect:idea}.  A description of the experimental
apparatus is given in section~\ref{sect:app}.  The results are
presented in section~\ref{sect:res}, with a discussion of these 
results in section~\ref{sec:disc}.  The feasibility of using an
improved version of this experiment to make a neutron lifetime
measurement is discussed in section \ref{sect:conclusion}.  The
techniques developed for this experiment have other potential
fundamental physics applications, also discussed in section
\ref{sect:conclusion}, such as improved sensitivity in the search for
the neutron's electric dipole moment\cite{Gol94} or detection of low
energy neutrinos\cite{McK00b}.

\section{Experimental Concepts}
\label{sect:idea}

\subsection{Magnetic Trapping}
\label{sec:trap}
Neutrons have a magnetic moment,
$\mathrm{\mu_{n} = -0.7~mK~T}^{-1}$, and interact with a
magnetic field, $\vec{B}$, by the dipole interaction:
\begin{equation}
    \label{eq:magint}
    H = -\vec{\mu}_{n}\cdot\vec{B} = |\mu_{n}|~\vec{\sigma}_{n} \cdot \vec{B},
\end{equation}
where the magnetic moment of the neutron is oriented anti-parallel to
its spin ($\sigma_{n}$).  When the neutron spin is aligned parallel to
the magnetic field the interaction energy is positive and regions of
high magnetic field repel neutrons in the parallel spin state.  These
neutrons are denoted ``low-field seekers''.

\begin{figure}[t]
    \begin{center}
	\includegraphics{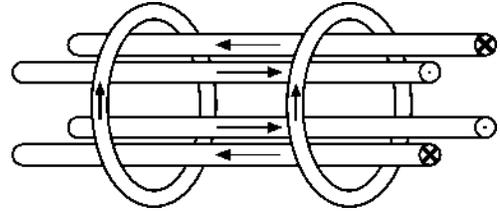}
    \end{center}
    \caption{A sketch of an Ioffe trap.}
    \label{fig:ioffe}
\end{figure}

Static magnetic traps create a magnetic field minimum in free space to
confine low-field-seeking particles.  One such trap, known as an Ioffe
trap, conceptually (and in our case schematically) consists of four wires
and two coils as shown in Fig.~\ref{fig:ioffe}.  This configuration
has been used to confine ions and plasmas and more recently to trap
neutral atoms\cite{Got62,Bag87}.  The four wires with alternating
current direction provide a cylindrical quadrupole field.  The two
coils, coaxial with the same current sense, provide a non-zero field
throughout the trapping region.  Our magnet is of this form and is
described further in section~\ref{sect:app}.

The interaction potential given in Eq.~\ref{eq:magint} also causes the
component of the magnetic moment perpendicular to the field to precess
around the local direction of the field with a frequency $\omega$
($\approx 180$~MHz at 1~T).  If this precession is considerably faster
than the change in the direction of the field seen by the neutron,
that is if
\begin{equation}
    \label{eq:adiab}
    \omega = \frac{2\mu_{n}B}{\hbar} \gg \frac{d\vec{B}/dt}{|\vec{B}|},
\end{equation}
then the neutron spin will remain aligned with $\vec{B}$ as it moves
throughout the trap (i.e.\ the neutron spin state will not change
relative to the field).  Under these conditions, one can essentially
view the magnetic potential, $H = |\mu_{n}| |B|$, as depending only on
the magnitude, not the direction, of the magnetic field.  Near regions
in which the field approaches zero this condition is not fulfilled and
trapped, low-field-seeking neutrons can invert their spin and be lost
from the trap (Majorana spin-flip losses).  Provided $|\vec{B}|$ is
sufficiently large (i.e.\ no zero-field regions) a neutron initially
in the trapped state will remain in the trapped state as it moves
throughout the trap.

\subsection{Superthermal Production of UCN}
\label{sec:stp}
To load neutrons into the trap their energy must be dissipated inside
the trapping region.  The ``superthermal'' process (so called because
it allows one to achieve phase space densities greater than that
coming from a ``thermal'' liquid hydrogen or deuterium cold source) is
an efficient method to load the trap\cite{Gol75,Gol77}.

\begin{figure}[t]
	\begin{center}
		 \includegraphics{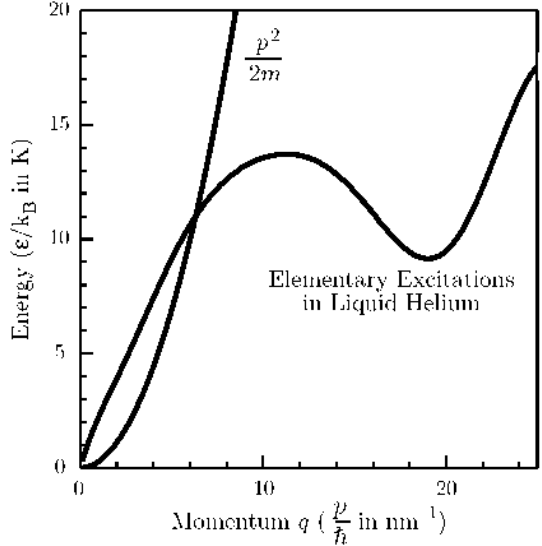}
	\end{center}
	\caption{The dispersion curve for excitations in superfluid
	helium and the free energy of the neutron\protect\cite{Cow71}.  The
	two curves cross at $q=0$ and at $q=q^{*}$, which corresponds
	to a neutron wavelength of 0.89~nm, or an energy of 12~K.}
	\label{fig:disp}
\end{figure}

The superthermal production of UCN in superfluid helium depends
critically on the fact that cold neutrons cannot scatter with
individual helium atoms as they would in a gas, but must scatter with
excitations (phonons and rotons) in the superfluid.  The dispersion
curve (energy versus momentum) for these excitations in superfluid
helium\cite{Cow71} is shown in Fig.~\ref{fig:disp} with the energy of
the free neutron ($p^{2}/2m$) overlaid for comparison.  For a neutron
to scatter to the vicinity of $q=0$ via the dominant single phonon
scattering process, momentum and energy conservation limit the initial
neutron momentum to values of $q\sim q^{*}$.  This results in the
creation of a single 12~K phonon in the helium.  Similarly, a UCN with
$E \sim 1$~mK can only ``upscatter'' by absorbing a 12~K phonon
(ignoring multi-phonon processes for the moment).  This fact allows
one to treat the neutron interaction with the superfluid as a two
level system with a 12~K energy difference between the two levels.

The one-phonon upscattering rate can be expressed as
\begin{eqnarray}
    \sigma(E\UCN\rightarrow E\UCN+\Delta) & = &\\ \nonumber
    & & \hspace{-1.2in} \frac{E\UCN+\Delta}{E\UCN}\/
    \exp(-\frac{\Delta}{k_{B}T})
    \sigma(E\UCN+\Delta \rightarrow E\UCN)
\end{eqnarray}
where $k_{B}$ is the Boltzmann constant.  For a fixed $\Delta$ (12~K
in our case), the upscattering rate can be made arbitrarily small
since it decreases exponentially with the temperature $T$ of the
helium bath.  Physically this corresponds to decreasing the population
of phonons of energy $\Delta$.  In practice, however, the rates of
other (multi-phonon) scattering processes dominate the upscattering
rate at low temperatures.  The second-order process is denoted as
two-phonon upscattering, although it applies equally to the scattering
of a neutron by a single phonon, in which the phonon is not absorbed. 
The rate of the two-phonon upscattering scales as $T^{7}$\cite{Gol79}
and has been measured down to 500~mK \cite{Gol83}.  In a helium bath
at a temperature of 1.16~K, the upscattering rate was $2.5\times
10^{-2}$~s$^{-1}$, corresponding to a storage time of 40~s.  At our
operating temperature of 250~mK, the upscattering rate should be 2000
times less than the beta-decay rate.  Minor modifications to the
apparatus now allow the helium to be cooled below 100~mK, making the
two-phonon upscattering rate negligible in performing a lifetime
measurement.

The use of superfluid helium to produce UCN has been demonstrated in
several previous experiments\cite{Gol83,Yos92,UCNbook,Kil87} and
upscattering rates have been measured in Refs.~\cite{Gol83} and
\cite{Kil87}.

\subsection{Detection of decays in the trap}
\label{sec:scint}
When a neutron decays in the trap, the recoiling electron scatters
from atomic electrons of the helium, ionizing tens of thousands of
helium atoms.  The electron deposits 300~keV~cm$^{-1}$ in the helium
with the range of a typical electron of order 1~cm\cite{Adams98}.  The 
helium ions tend to form diatomic molecular ions before
recombining with electrons, leading to the creation of excited-state
diatomic helium molecules.  These molecules are created in both
singlet and triplet states.

When each helium molecule decays it emits a photon in the extreme
ultraviolet (EUV; $\lambda \approx 70-90$~nm).  The singlet decays
give a relatively intense pulse of light in a short time; 4000~EUV
photons in less than 20~ns for a typical neutron decay\cite{Adams98}. 
These singlet decays and are used to detect the neutrons decaying in
the trap.  The longer decay time for the triplet states renders them
less useful for this purpose.  The same number of photons emitted over
13~s\cite{McK99} makes it impossible to distinguish the triplet events
from backgrounds.

To detect the light pulse emerging from the singlet decays directly in
the trap would require EUV sensitive detectors that work at low
temperatures ($< 250$~mK), in liquid helium, and inside an intense
magnetic field (1~T).  The detectors would also be exposed to the cold
neutron beam, or at least to neutrons scattering in the helium, and
would need to tolerate this radiation without creating backgrounds. 
Finally, there is limited space available inside the magnet bore. 
Since it is extremely difficult to satisfy these detector
requirements, the scintillation light is detected indirectly.

The EUV scintillation light is converted to blue/violet light inside
the apparatus using a pure hydrocarbon, 1-1-4-4 tetraphenyl 1-3
butadiene (TPB).  TPB absorbs EUV light and emits light in a spectrum
ranging from 400~nm to 500~nm\cite{Bur73}.  This light can then be
transported out of the trapping region and into detectors at room
temperature.  The detector insert is described in detail in
section~\ref{sect:app}.

\subsection{Experimental Method}
\label{sec:oview}
Each experimental run consists of a loading phase in which the
neutrons are loaded into the trap and an observation phase in which
the trapped neutrons are detected as they decay.  At the start of a
run the magnetic field is on forming the magnetic trap which will
confine the neutrons.  The neutron beam is turned on and neutrons
begin to scatter in the liquid helium, producing UCN which become
confined in the trap.  The UCN creation rate is proportional to the
incident neutron flux and is constant during the loading phase.  UCN are
lost from the trap by beta decay in proportion to the number trapped. 
Thus, neutrons accumulate in the trap with a time constant of the
neutron lifetime or approximately 15~min.  In the actual experiment,
the trap is loaded for 22.5~min, yielding 78~\% of the trapped UCN
population that could be achieved with an infinite loading time.  Once
the beam is turned off, the number of neutrons in the trap decreases
due to beta decay.  The photomultiplier tubes are turned on 10~s after
the beam is turned off and neutron decays are detected for 1~h, giving
a time-varying signal directly proportional to the decreasing number
of neutrons in the trap.  Trapped neutrons are indicated by an
exponentially decaying signal consistent with the neutron beta-decay
lifetime, or a more rapid decay if there are additional trap losses.

Two classes of problems could prevent observation of this signal:
spurious losses (which cause the initial trapped population to be
significantly lower due to losses during loading) and background
events (which could give a much higher overall counting rate and
obscure a trapped neutron signal).  The next two subsections will
treat these two issues.

\subsection{Loss Mechanisms}
\label{sec:traploss}
Known loss mechanisms include one- and two-phonon upscattering (which
as discussed above is reduced by cooling the helium target), Majorana
spin-flips (which are minimized by the non-zero-field character of the
trap), and $^{3}$He absorption.  In addition, some neutrons with
energies greater than the trap depth (that is, sufficient to allow
them to escape) may be confined by the field for an extended period of
time but escape from the trap before they decay.

Helium has two stable isotopes, $^{3}$He and $^{4}$He.  While $^{4}$He
does not absorb neutrons, $^{3}$He has an extremely high absorption
cross-section.  For UCN contained within a bath of
$^{4}$He with a small amount of $^{3}$He present, the absorption rate
is given by:
\begin{equation}
    \Gamma_{abs} = x_{3} \cdot n \cdot \sigma_{th} \cdot v_{300\rm K} 
                 = x_{3} \cdot 2.6\times 10^{7}~\mathrm{s}^{-1}
\end{equation}
where $x_{3}$ is the ratio of $^{3}$He atoms to $^{4}$He atoms, $n$ is
the density of helium nuclei, $\sigma_{th}$ is the thermal cross
section for neutron capture onto $^{3}$He and $v_{300\rm K}$ is the
neutron velocity at room temperature.  Isotopically purified $^{4}$He
is available with $x_{3} < 5\times 10^{-16}$\cite{Mcc78}.  At this
limit, the absorption rate is $1.3\times 10^{-8}\mathrm{~s}^{-1}$, or
ten thousand times less than the beta-decay rate.  This gives a
negligible shift in the trap lifetime.  ``Reversing'' this analysis
allows the data presented in section~\ref{sect:res} to yield a three
standard deviation upper limit on the $^{3}$He concentration of $x_{3}
\le 5\times 10^{-11}$.  A more accurate, independent measurement of
$x_{3}$ using accelerator mass spectroscopy is presently being
pursued.

A ``marginally trapped'' neutron is one with an energy greater than
the trap depth, but in a trajectory that keeps the neutron confined in
the trapping region for a significant time.  Simulations have
determined that the vast majority of these neutrons will be expelled
from the trap within a few seconds\cite{Carlo}.  In addition, it has
been shown through analytical calculations\cite{Bro00} that by
lowering the magnetic field to 30~\% of its original value and then
raising it back up, a trapped sample is achieved that has no neutrons
stored with energies above the trap depth and about one-half of the
original trapped neutrons.  For this demonstration of trapping, this
procedure is not necessary.

\subsection{Backgrounds}
\label{sec:bak}
Background events are any events recorded by the data acquisition
system that arise from sources other than decaying neutrons.  The data
acquisition system records an event when coincident PMT pulses above a
set charge threshold occur on the two photomultipliers observing the
trapping region.  Events which are coincident with the detection of
light pulses from PMTs observing plastic scintillator paddles placed
below the primary detection cell are vetoed.

Hence the detection system is designed to reject several expected
backgrounds.  Cosmic ray muons pass through materials at a rate of
1~cm$^{-2}$~min$^{-1}$\cite{PDG} and penetrate any reasonable
shielding.  They leave an ionization trail as they pass through,
depositing roughly 2~MeV~g$^{-1}$cm$^{2}$ times the density of the
material\cite{PDG}.  Pulses in the primary detection system due to
cosmic ray muons are vetoed by the auxiliary paddles and do not
produce background events.

The requirement of coincidence between two PMTs observing the trapping
region suppresses background events caused by neutron induced
luminescence of materials.  The neutron absorbers used to shield the
materials surrounding the trapping region from the neutrons (primarily
hexagonal boron nitride, hBN) glow after exposure to neutrons with a
magnitude decreasing slowly on a time scale comparable to an
experimental run\cite{Huf01}.  This luminescence (which was for some
time the major obstacle to extracting the trapped neutron signal) was
attenuated by placing graphite between the hBN and the light
collection system and by replacing the collimator and beam stop with
B$_{4}$C (a material which has considerably less luminescence).  To
suppress this background, coincident detection in two PMTs is
employed.  A signal is recorded when at least two photons are detected
in each PMT within a 23~ns window.  With this threshold, the
efficiency for detecting neutron decays is 31~\%.

The system is also susceptible to background events from other
sources.  The passage of high-energy particles through any part of the
trapping region or detection system is expected to cause
scintillations in the primary PMTs.  Other than cosmic ray muons
which are vetoed, any particle that causes a detected scintillation
will produce a background event.  Such particles include decay
products from radioactive nuclei, either naturally occurring or
created by neutron capture during the loading phase, and neutral
particles (gamma rays or fast neutrons) created outside of the
apparatus.  Unlike muons, gamma rays and fast neutrons tend to lose a
significant fraction of their energy in each scattering event and
cannot be detected with veto detectors.  To reduce background events
from gamma rays, the apparatus is shielded with 10~cm of lead.

Backgrounds from decays of isotopes created by neutron absorption
decay exponentially with their respective lifetimes.  To minimize such
backgrounds, the materials exposed to the neutron beam are carefully
chosen such that their primary constituents have small neutron
absorption cross sections (or absorb neutrons to form a stable
isotope, thus giving no background).

Despite several successful methods of reducing backgrounds, a decaying
signal with peak magnitude of roughly 0.2~s$^{-1}$ had to be extracted
from a time-varying background signal of about 6~s$^{-1}$ initially,
with a constant component of 2~s$^{-1}$ at longer times.  This was
accomplished by measuring the background directly, taking half of the
runs with the magnetic trap initially turned off, and subtracting this
background from the trapping runs.

\section{Experimental Apparatus}
\label{sect:app}
\subsection{Beam}
The experiment resides at the end of cold neutron guide number six
(NG-6) in the guide hall of the NIST Center for Neutron Research
(NCNR)\cite{NCNR}.  Neutrons are released from the 20~MW reactor and
are moderated by D$_{2}$O surrounding the reactor core and then by a
liquid hydrogen cold source operating at a temperature of 20~K.
Neutrons partially thermalize with the hydrogen and exit towards the
neutron guides.  Some fraction of these neutrons pass into the open
end of NG-6, a rectangular $^{58}$Ni-coated guide 60~m long, 15~cm
tall and 6~cm wide.

Neutrons exiting the far end of NG-6 pass through a 6~cm diameter
collimating aperture ($C_1$) and then through filtering materials
consisting of 10~cm of bismuth and 10~cm of beryllium.  The single
crystal bismuth attenuates the flux of gamma rays from upstream
sources.  The polycrystalline blocks of beryllium Bragg-reflect
neutrons with wavelengths less than 0.395~nm, removing them from the
beam.  The neutron beam exiting the filter has a measured capture flux
of $9\times 10^{8}$~n~cm$^{-2}$s$^{-1}$\cite{nico}.

Immediately preceding the trapping region, a second aperture ($C_2$)
collimates the beam before it enters the trapping region.  The
collimators $C_1$ and $C_2$ are chosen such that any neutron passing
through both collimators will be absorbed in the beamstop at the end
of the trapping region unless it scatters in the helium.

\begin{figure}[t]
    \begin{center}
	\includegraphics{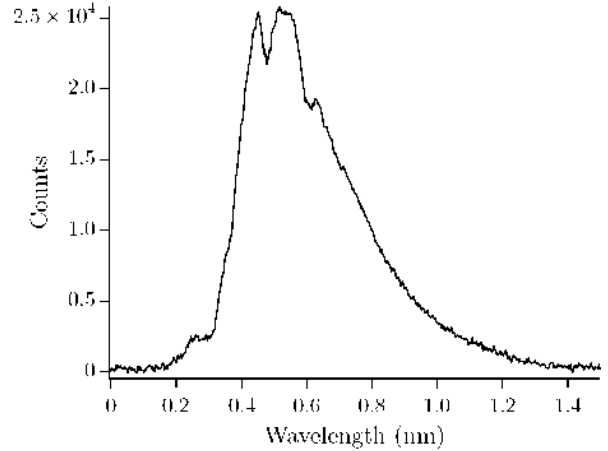}
    \end{center}
    \caption{The neutron spectrum measured at the end of NG-6 using 
	    time-of-flight techniques.}
    \label{fig:TOF_dat}
\end{figure}

The wavelength spectrum of the beam is measured in a separate
experiment by the time-of-flight method.  The beam is collimated to a
diameter of 400~$\mu$m. A cadmium disc positioned directly behind the 
collimator and rotating at 67~Hz
blocks the beam except when either of a pair of slits 250~$\mu$m wide
passes in front of the collimator.  Neutrons passing through the slit
are detected using a $^{3}$He detector placed 69~cm behind the
chopper.  The output of the $^{3}$He detector is recorded with a
multi-channel scaler, with a 5~$\mu$s bin width.  The scaler is
triggered using a light emitting diode (LED) and a photodiode
detecting the passage of the slit approximately 180$^{\circ}$ from the
neutron beam.  Data from the time-of-flight spectrum is shown in
Fig.~\ref{fig:TOF_dat}.

Two points in the spectrum are available to set the initial time, the
beryllium edge, at 0.395~nm, and the narrow slice of the beam removed
by an upstream monochromator in the neutron guide at 0.48~nm.  Both
the unnormalized capture flux and 0.89~nm flux (important for UCN
production using scattering from liquid helium) can be extracted. 
Using the known capture flux for normalization, the flux at 0.89~nm is
\begin{equation}
   \label{eqn:this}
    \frac{d\Phi}{d\lambda}= (1.62\pm 0.08)\times 10^{6}
	   \mathrm{~n~cm^{-2}~s^{-1}~K^{-1}}.	   
\end{equation}
The error bars are one standard deviation and primarily correspond
to uncertainties in the wavelength calibration.

\subsection{Magnet}
\label{sec:mag}
\begin{figure}[t]
    \begin{center}
        \includegraphics{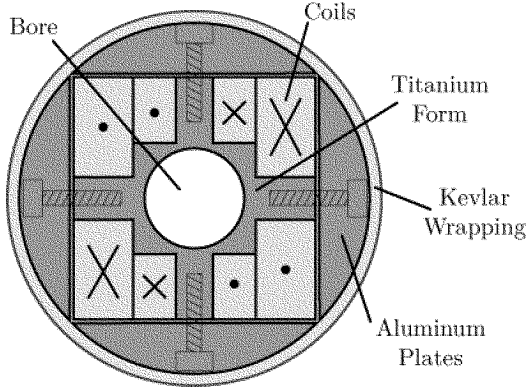}
    \end{center}
    \caption{Cross-sectional view of the quadrupole magnet form.}
    \label{fig:form}
\end{figure}

The UCN are confined in an Ioffe-type superconducting magnetic
trap\cite{Got62}.  The magnet uses four racetrack-shaped coils to
produce the quadrupole field.  These consist of two different size
coils and are assembled as shown in Fig.~\ref{fig:form}.  The two
larger coils, with 777 turns each, are 69.1~cm long by 9.2~cm high,
with a coil cross-section of 3.2~cm by 2.0~cm.  The smaller coils,
with 336 turns each, are 65.2~cm long by 5.3~cm high, with a cross
section of 1.5~cm by 2.0~cm.
The coils are placed onto the 6-4 ELI grade titanium form and isolated
using Kapton\footnote{Certain commercial equipment, instruments, or
materials are identified in this paper in order to specify the
experimental procedure adequately.  Such identification is not
intended to imply recommendation or endorsement by the National
Institute of Standards and Technology, nor is it intended to imply
that the materials or equipment identified are necessarily the best
available for the purpose.} and G-10 spacers.  Aluminum compression
plates, machined to give a cylindrical outer surface, are attached to
the form to provide compression of the coils.  This cylindrical
assembly is wrapped under 67~N of tension with 20 layers of Kevlar to
uniformly compress (prestress) the coils.  To protect the Kevlar, the
magnet was first wrapped in a layer of Tedlar then the Kevlar, and
then an outer layer of fiberglass with the wrappings embedded in
epoxy.  The quadrupole assembly had an outer diameter of 14.5~cm and
an inner bore of 5.08~cm.  The ends of the titanium form were machined
to mate directly with the inner vacuum can of the cryostat.

The solenoid assembly surrounds the quadrupole assembly and consists
of two sets of solenoids.  Each set contains a primary solenoid
which is flanked by a pair of smaller solenoids with opposite
polarity.  These cause the magnetic field of the primary solenoid to
fall off more quickly with distance along its axis.  The primary
solenoids each consist of 1564~turns with an inner diameter of
14.7~cm, outer diameter of 18.2~cm and length of 5.15~cm.  The two
``bucking'' coils placed 5.0~cm on each side of the primary coil are
365~turns with an inner diameter of 14.7~cm, outer diameter of 21.5~cm
and length of 1.15~cm.

\begin{figure}[t]
    \begin{center}
	\includegraphics{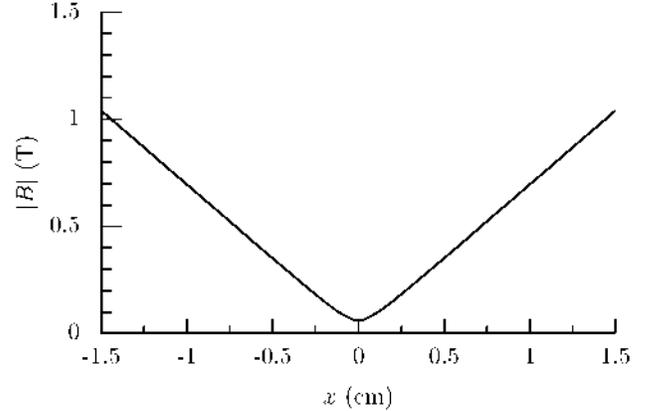}
    \end{center}
    \caption{The magnetic field along the \textit{x}-axis ($y=z=0$).}
    \label{fig:quadprof}
\end{figure}

\begin{figure}[tb]
    \centering
    \includegraphics{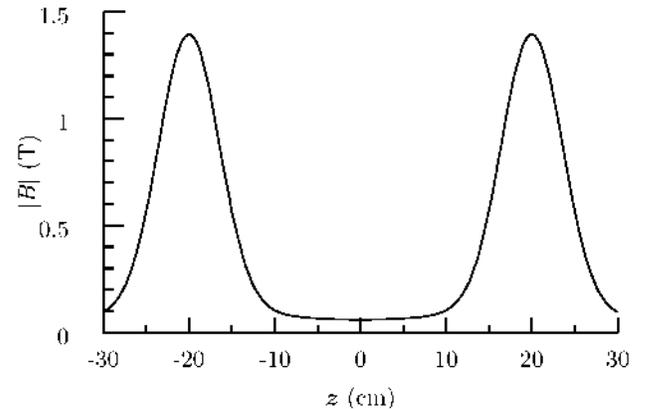} 
    \caption{The magnetic field along the \textit{z}-axis ($x=y=0$),
    showing the ``pinch'' resulting from each solenoid, closing the
    ends of the trap.}
    \label{fig:solprof}
\end{figure}

The magnitude of the quadrupole field increases linearly with radius
with a gradient of 0.69~T~cm$^{-1}$ (at 180~A).  Along the
axis of the form, the field minimum (0.11~T at 180~A) is provided by
the solenoids.  Denoting this axis as the $z$-axis, the magnitude of
the field along the $x$-axis is shown at $y=z=0$ in
Fig.~\ref{fig:quadprof}.
The solenoids each create a short high-field region closing off the
ends of the trapping region.  The field along
the $z$-axis is shown at $x=y=0$ in Fig.~\ref{fig:solprof}.

The depth of the magnetic trap is defined as the difference between
the lowest field point at which neutrons can be lost from the trap
($B_{max}$) and the lowest field point in the trap ($B_{min}$).  In
our geometry, the solenoids were chosen to produce a peak field higher
than the walls of the quadrupole.  The trap depth is thus set by the
quadrupole field at the inner edge of the detector insert (see below). 
This occurs at a radius of 1.59~cm, corresponding to a trapping field
of 1.15~T\@.  Subtracting the minimum field of 0.11~T gives a net trap
depth of $B_{T}=1.04$~T\@.

\subsection{Cryogenics}
\label{sec:fridge}
\begin{figure}[t]
	 \begin{center}
		 \includegraphics{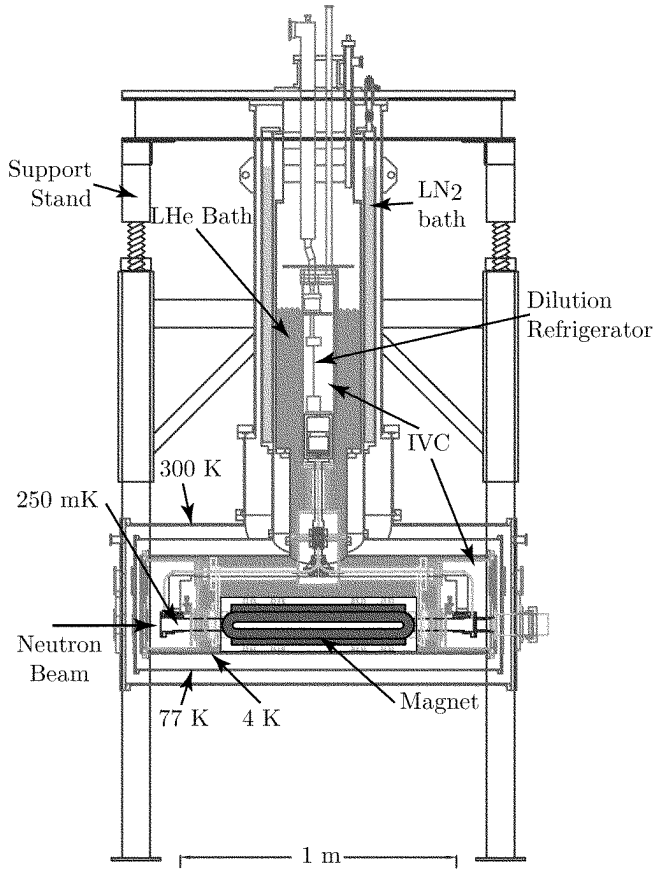}
	 \end{center}
	 \caption{A sketch of the T-shaped dewar.}
	\label{fig:dewar}
\end{figure}

Cooling the helium target below 250~mK requires the use of a
$^{3}$He-$^{4}$He dilution refrigerator which is connected to
horizontally oriented magnetic trapping region.  This required a
custom-designed cryogenic dewar, consisting of a horizontal section to
house the magnet and trapping region and a vertical section to house
the refrigerator and is shown in Fig.~\ref{fig:dewar}.

The neutron beam enters from one end of the dewar and light from the
neutron decays exits from the opposite end.  The neutron entrance
windows consist of two layers of different materials, one to make a
vacuum seal and the other to block black-body radiation, with both
having good neutron transmission properties.  The vacuum windows are
made from a clear fluoropolymer, Teflon PFA. A simple compression
seal, where the edges of a Teflon disc are compressed between two
metal flanges until the Teflon begins to deform, provides the vacuum
seal\cite{But98a}.  The windows used at 250~mK (where black-body
absorbers are not needed) are 250~$\mu$m thick, 6.35~cm in diameter,
leaving a 6.35~mm wide section compressed around the circumference of
a window 5.08~cm in diameter.  The other two windows at 300~K and 4~K
are similar with a 7.62~cm diameter disc making a window 6.35~cm in
diameter.  To block blackbody radiation, a 50~$\mu$m thick, 6.35~cm
diameter beryllium disc (Grade PF-60, Brush Wellman) covered the
Teflon window at both the 4~K and 77~K end flanges.  The light
transmission windows, on the opposite end of the cryostat, are made
from ultra-violet transmitting (UVT) acrylic (polymethyl methacrylate
or PMMA) and are sealed using epoxy, indium o-ring seals, or rubber
o-rings.

The isotopically pure $^{4}$He and detector insert are contained
within the inner vacuum can (IVC) and held fixed with respect to the
magnet and neutron beam, while thermally anchored to the dilution
refrigerator.  The helium is contained within the IVC in a tube made
of 70-30 cupronickel which extends through the magnet and into the end
can of the dewar.  There, a copper end cap is soft-soldered (using
50/50 tin/lead solder with Nokorode paste flux) to each end of the
tube to provide a sealing surface for either the Teflon window or the
light transmission window.  The endcaps are thermally anchored to the
copper heat link which extends to the dilution refrigerator.

The cell is filled with isotopically purified $^{4}$He by condensing
it in from room temperature.  The helium is stored as a low pressure
room temperature in two 450~L electropolished stainless steel tanks. 
The storage volume is connected to the cell through stainless steel
tubing and all-metal valves.  Two independent fill lines enter the
dewar and each are thermally anchored at various points of the
dilution refrigerator before they connect to the cell.  The heat link
and the fill lines are flexible near the T-section.  The cell and
lower half of the heat link is supported directly from the IVC using
Kevlar braid.

\subsection{Neutron Shielding}
\label{sec:BN}
When the neutron beam is open to the trapping region, $4\times
10^{8}$~n~s$^{-1}$ enter the helium bath.  At the end of the trapping
region, those neutrons which have not scattered in the helium are
absorbed in the beam dump.  Since the scattering cross section for
cold neutrons in liquid helium\cite{Som55} decreases rapidly with
wavelength, at 0.89~nm, only 6~\% of the neutrons scatter, while at
0.45~nm, almost 40~\% scatter.  As a result, about 2$\times
10^{8}$~n~s$^{-1}$ (or a total of 3$\times 10^{11}$ neutrons) scatter
into the walls when the beam is on, potentially producing backgrounds.

\begin{figure}[t]
    \begin{center}
	\includegraphics{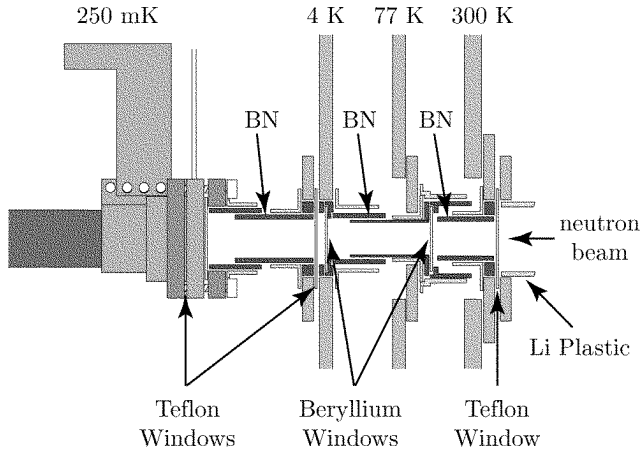}
    \end{center}
    \caption{Sketch of the windows in the dewar end flanges
    on the beam entry end showing the interlocking boron nitride
    tubes shielding the beam.}
    \label{fig:efbeam}
\end{figure}

To minimize such backgrounds, neutron absorbing materials are placed
such that scattered neutrons do not reach materials that can activate. 
Several different materials are used, all of which contain either
lithium or boron, materials which have high neutron absorption cross
sections.  A sketch of the shielding and windows on the beam entrance
end of the dewar is shown in Fig.~\ref{fig:efbeam}.

The sensitivity to activation-induced backgrounds is greatest inside
the detector insert and lightpipes.  To minimize activation in the
detection system, the materials inside the shielding are carefully
chosen to contain elements which have minimal activation such as
hydrogen, carbon and oxygen.  Surrounding these materials is a shield
of hexagonal boron nitride (hBN, grade AX05, Carborundum Corporation)
to absorb the remaining neutrons.  This shield consists of a series of
interlocking hBN cylinders (5~cm long, with an outer diameter of
4.15~cm and a wall thickness of 1.5~mm).  hBN has a neutron
attenuation length of 56~$\mu$m at 0.89~nm.

The hBN is coated with a thin layer of colloidal graphite (Aerodag G,
Acheson Industries) to attenuate the luminescence (see below).  In
addition, non-interlocking tubes of graphite (15~cm long with a wall
thickness of 0.2~mm, Grade AXF-5Q1, Poco Graphite) are inserted
between the boron nitride tube and detection region.

\subsection{Detection System}
\label{sec:tpb}
\begin{figure}[t]
	\begin{center}
 		\includegraphics{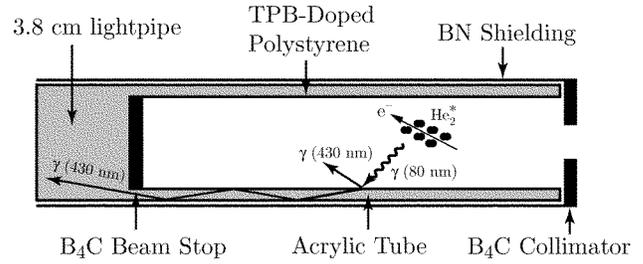}
	\end{center}
	\caption{A sketch of the detector insert with the scintillation
		process and light transport shown (Not to scale).}
	\label{fig:detection}
\end{figure}

The detector insert is a UVT acrylic tube whose inner surface is
coated with TPB-doped polystyrene (mass fraction of 40~\% TPB). 
This insert, shown in Fig.~\ref{fig:detection}, together with a series
of acrylic windows and lightpipes leading to a pair of PMTs, forms a
detection system allowing neutron decay events to be recorded as a
function of time.

\begin{figure}[t]
    \begin{center}
	\includegraphics{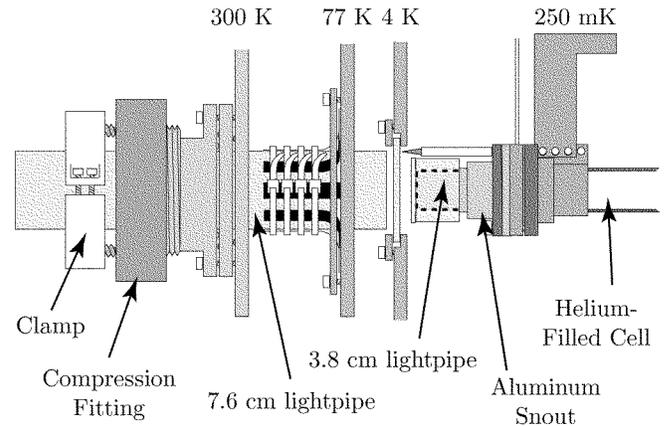}
    \end{center}
    \caption{Sketch of the windows in the dewar end flanges,
	on the light exit end.}
    \label{fig:eflight}
\end{figure}

Each beta-decay event in the trapping region gives rise to a pulse of
EUV light in the helium.  This light is downconverted to the visible
by the TPB and transported past the opaque B$_{4}$C beamstop by the
acrylic tube (3.8~cm outer diameter, 3.1~mm wall thickness) to a
3.8~cm diameter solid acrylic lightpipe.  This lightpipe extends to
the end of the helium-filled cell (see Fig.~\ref{fig:eflight}).  At
the end of the cell, the light passes through a pair of UVT acrylic
windows into a 7.6~cm diameter lightpipe which extends out of the
dewar.  The window on the cell is formed by epoxying (Stycast 1266) an
acrylic window to an acrylic tube.  The other end of the tube is
epoxied to an thin (0.5~mm wall thickness) aluminum cylinder
(``snout'') which attaches to the cell body.  The 3.8~cm lightpipe is
epoxied to the window of the snout to prevent the detection insert
from moving with respect to the cell.  The 7.6~cm lightguide is
heatsunk to the 77~K radiation shield using strips of aluminum foil as
shown in the figure. 

\begin{figure}[t]
	\begin{center}
		\includegraphics{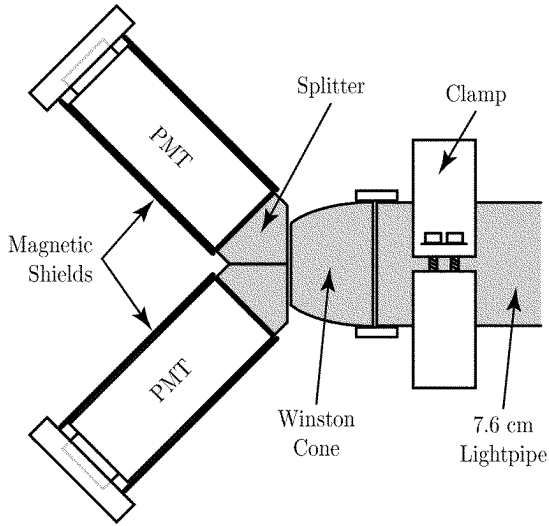}
	\end{center}
	\caption{A sketch of the optical elements connecting the
	    7.6~cm lightpipe to the PMTs.}
	\label{fig:pmts}
\end{figure}

Outside the dewar, the light emerging from the 7.6~cm lightpipe is
coupled into a pair of 5~cm diameter PMTs using a series of optical
elements (Fig.~\ref{fig:pmts}).  The light is first ``condensed'' as
it passes through the Winston cone\cite{Win70}.  The light exiting the
Winston cone is split into two 5~cm diameter lightpipes using a
splitter which is made from two 5~cm diameter lightpipes machined at a
$45^{\circ}$-angle and epoxied together, forming an ell.  The corner
is machined away leaving a flat elliptical surface with a 5~cm minor
axis and a 7.1~cm major axis which is attached to the Winston cone. 
The Winston cone and the splitter are diamond turned, polished, and
aluminized to maximize the light transmission.  The assembly has an
80~\% transmission efficiency from the 7.6~cm diameter lightpipe to
the two ends of the splitter.  All of the optical elements are
contained within an aluminum housing which both blocks light and
creates a space through which dry N$_{2}$ gas is flowed to prevent
helium contamination of the PMTs.

The photomultipliers are Burle model 8850.  The signal is read out at
the anode as a discrete current pulse with the tubes biased using
positive high voltage.

\begin{figure}[t]
	\begin{center}
		\includegraphics{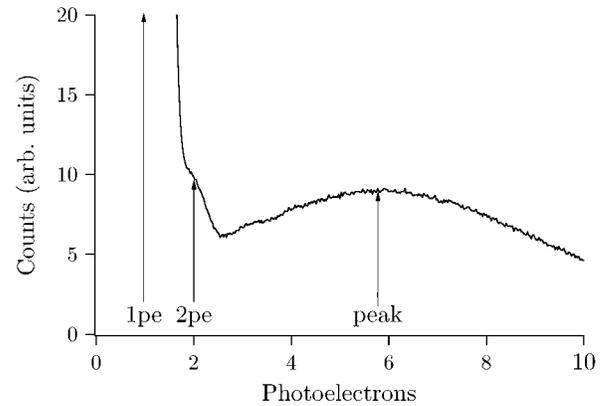}
	\end{center}
	\caption{Pulse area spectrum from 360~keV beta line spectrum.}
	\label{fig:calib}
\end{figure}

A series of calibrations was performed using both alpha and beta
sources to determine the efficiency for detecting neutron decay
events.  Using a $^{113}$Sn beta line conversion source
($E_{\beta}=360$~keV), the assembled detector insert and light
collection system were tested using a single PMT at the end of the
Winston cone.  The resulting spectrum is shown in
Fig.~\ref{fig:calib}, with a typical pulse on average giving between
5.5 and 6.0 photoelectron signal.  A series of tests using the higher
energy $^{210}$Po alpha source ($E_{\alpha}=5.3$~MeV) gives the
relative detection efficiency for EUV light produced at different
longitudinal positions along the cell.  The efficiency varies from
75~\% to 125~\% (relative to the efficiency at the center of the cell)
along the length and has an approximately linear dependence.

A simple Monte Carlo is used to calculate our efficiency for detecting
trapped neutrons.  Each event is modeled by taking a random position
along the length of the cell and a random beta energy, weighted by the
known shape of the beta spectrum.  The results of the $^{113}$Sn
calibration give an output of $14.5 \pm 1.5$ photoelectrons per MeV
of electron kinetic energy.  The expected signal is corrected for the
measured 80~\% efficiency of the splitter and for the dependence of
the signal on the position of the event in the cell.  Requiring
coincidence between pulses in both PMTs, each having an integrated
area corresponding to two or more photoelectrons, results in the
detection of $31~\% \pm 4$~\% of neutron beta decays in the trapping
region.

\subsection{Data Acquisition}
\label{sec:daq}

\begin{figure}[t]
	\begin{center}
		 \includegraphics{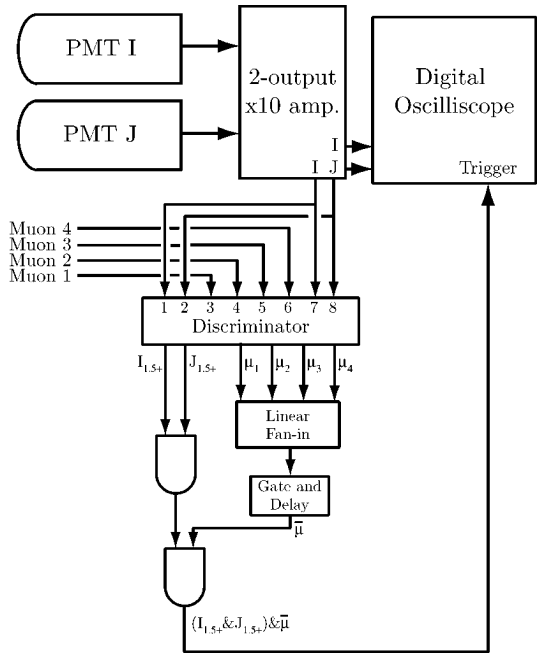}
	\end{center}
	\caption{Data acquisition schematic}
	\label{fig:DAQ}
\end{figure}

A schematic of the data acquisition system is shown in
Fig.~\ref{fig:DAQ}.  The anode signal from each of the two PMTs
looking at the decay region (labeled I and J) is fanned into three
outputs.  One output connects directly to a digital oscilloscope,
while the other two, along with the four muon veto detectors, pass
into a discriminator.  The voltage thresholds for the PMT signals are
set to trigger on each pulse which would give a charge-integrated
signal of two or more photoelectrons (denoted as I$_{1.5+}$ and
J$_{1.5+}$), as determined empirically.  The muons are detected using
three plastic scintillators ($\rm 2.5~cm \times 32~cm \times 1.42~m$)
which surround the lower half of the dewar and a fourth detector with
slightly smaller dimensions that is placed over the PMTs and
connecting lightguides.  The muon thresholds are set such that each
detector produces about 1000~s$^{-1}$.  Varying this threshold so as
to change this counting rate by more than a factor of two had no
effect on the 8~s$^{-1}$ muon coincidence rate with the detector
insert, indicating that all true coincidences are being included at
these threshold levels.

The outputs I$_{1.5+}$ and J$_{1.5+}$ are ANDed to produce a
coincident I$_{1.5+} \cdot$ J$_{1.5+}$ signal 23~ns long.  The
inverted output of the 1.5+ coincidence signal is delayed and ORed
with the delayed combined muon signal to produce $\overline{(
\overline{(I_{1.5+} \cdot J_{1.5+})} + \mu )} = (I_{1.5+} \cdot
J_{1.5+}) \cdot \overline{\mu}$, or a coincidence between I$_{1.5+}$
and J$_{1.5+}$, where no muon paddle has recorded a coincident muon. 
This signal is used to trigger the digital oscilloscope.

In addition to the above, the trigger output signal from the
oscilloscope is recorded.  This signal is used in combination with the
number of times that the coincidence circuit tried to trigger the
oscilloscope to correct for the deadtime of the oscilloscope and to
assign time information to each scope event.  The counters are active
for 98~\% of the time, with a dead time of 20~ms during the read out
by the computer.

Digitizing the pulses with an oscilloscope allows one to integrate the
area of each PMT pulse and make a cut on the integrated pulse area
instead of the peak voltage.  The integrated charge gives a sharper
distinction between one- and two-photoelectron events, increasing the
signal-to-background discrimination efficiency.

The first two input channels of the oscilloscope were connected to the
amplified outputs of the two photomultiplier tubes I and J\@.  The
third, used to trigger the oscilloscope, is connected to the logic
signal $(I_{1.5+} \cdot J_{1.5+}) \cdot \overline{\mu}$.  For each
event, the oscilloscope recorded data at 250 megasamples per second
for 1~$\mu$s.  After 100 trigger events were recorded, the scope was
read out by the computer, giving a deadtime of nearly one second
between sets of 100 triggers.

\section{Results}
\label{sect:res}
\begin{figure}[t]
	\begin{center}
		\includegraphics{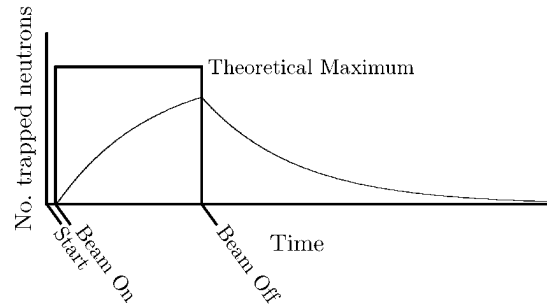}
	\end{center}
	\caption{A plot of the predicted trapped neutron population
	over the course of a data run.  Three times are indicated: the
	start of the run ($t=0$~s) and the times at which the neutron
	beam is turned on ($t=100$~s) and off ($t=1450$~s).  The run
	ends at $t=5050$~s.}
	\label{fig:posplot}
\end{figure}

Data were taken in a series of runs, each divided into three phases
(Fig.~\ref{fig:posplot}).  For the first 100~s, the neutron beam is
off.  This data are used for diagnostic purposes to detect shifts in
the background event rate (such as from neutron activation of
materials in the trapping region with long lifetimes).  In the loading
phase, for the next 1350~s, the neutron beam is on and the trapped UCN
population slowly increases, asymptotically approaching the
theoretical maximum.  During this phase the photomultiplier tubes are
turned off and no events are recorded.  In the last, or observation
phase, data are recorded as the trapped neutron population decays for
3600~s through beta decay.  Data are recorded both in the form of
scaler totals each second and sets of one hundred oscilloscope events.

Two types of runs were performed.  In a ``trap-on'' run the magnetic
trap is energized throughout the entire run.  In a ``trap-off'' run
the magnet is initially off but is brought up to full current during
the last 50~s of the loading phase (making conditions during the
observation, phase identical between trap-on and trap-off runs).  This
results in a small number of trapped UCN in the trap-off runs,
estimated to be at most 2~\% of that in the trap-on runs.  The
non-trapping or trap-off data are subtracted from the trap-on data.

To verify that the difference in signal observed between trap-on and
trap-off runs is due to the decay of trapped UCN and not from
backgrounds, additional runs were necessary.  Trap-on and trap-off data
were taken in the same manner but with other conditions changed so
that no UCN should be trapped.  Specifically, data are taken with the
temperature of the helium bath at 1.2~K, rather than at 250~mK
(denoted as ``warm runs'') and separately with the helium bath doped
with enough $^{3}$He ($x_{3}=2 \times 10^{-7}$) to quickly absorb
trapped UCN without significantly attenuating the neutron beam
(denoted as ``$^{3}$He runs'').

The trapping data were acquired over the course of two reactor cycles. 
During the first reactor cycle, from 3/23/99 to 4/5/99, a series of
trapping runs was taken, followed by a set of warm runs, more trapping
runs, and finally a set of $^{3}$He runs.  Between reactor cycles the
cell was pumped out and flushed several times with isotopically pure
helium in order to cleanse the cell of residual $^{3}$He from the
$^{3}$He runs.  During the next reactor cycle, from 4/16/99 to
5/23/99, a longer series (about 18~days) of trapping runs was taken,
followed by a series (about 20~days) of $^{3}$He runs.

In the first reactor cycle, the magnet operated at a current of 180~A,
giving a trap depth of 0.75~mK\@.  During the second cycle, a
grounding problem developed in the magnet and it operated at a lower
current of 120~A, resulting in a trap depth of only 0.5~mK\@.  The
number of trapped neutrons theoretically scales with magnetic field as
$B^{3/2}$, resulting in the number of trapped neutrons being a factor
of $\approx 0.5$ smaller than in the first cycle.  The background
count rate remained the same, thus even with four times as many runs,
the signal extracted from the second set of data is statistically
equal to that from the first set.

353 pairs (trap-on and trap-off) of data runs were taken, 87 with a
trap depth of 0.75~mK and 266 with a trap depth of 0.5~mK. For the
0.75~mK set, there are 32 pairs of trapping runs, 31 pairs of warm
runs, and 24 pairs of $^{3}$He runs.  In the 0.5~mK set, there are 128
pairs of trapping runs and 136 pairs of $^{3}$He runs.

Data are collected using two parallel systems: storage of events using
an array of scalers and digitization of the event pulses using a
digital oscilloscope as discussed above.  The oscilloscope data is
binned into 1~s bins and is corrected for deadtime during the
oscilloscope readout.  This deadtime is strongly dependent on the
trigger rate, which is decreasing in time over the course of a run. 
The exact bin in the scaler data in which the oscilloscope is being
read is known.  Any counts which occur in these bins are thrown out
and the entire one second bin is declared as deadtime to avoid any
bias.  The data are then histogrammed into bins of fifty seconds where
the deadtime is small and can be corrected by scaling the data by one
over the live fraction.  Data from the first 50~s of observation in
each run contain too much deadtime to be accurately corrected in this
manner and instead are removed.

The data of a given type (trapping, warm or $^{3}$He; trap-on or
trap-off; 0.75~mK or 0.5~mK) are pooled together in fifty second bins
and deadtime corrected.  These sets of data are compared, subtracting
the pooled trap-off data from the pooled trap-on data, to look for a
signal (or lack thereof) indicating the decay of trapped neutrons.

\begin{figure}[tb]
	\begin{center}
		\includegraphics{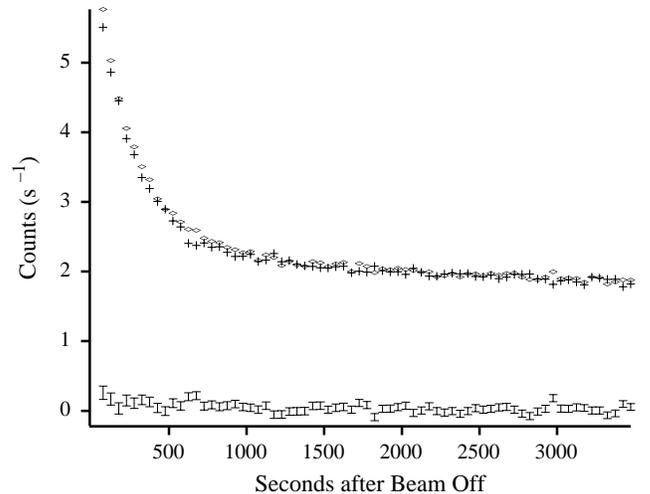}
	\end{center}
	\caption
	{A plot of the neutron trapping data from the first reactor cycle,
	with the average of the pooled trap-on runs and the pooled
	trap-off runs shown separately.  The difference is shown below,
	with statistical error bars.}
	\label{fig:cycle1pn}
\end{figure}

The pooled data from the first reactor cycle are shown in
Fig.~\ref{fig:cycle1pn}.  The average of the 32 trap-on runs and the
32 trap-off runs is shown starting 50~s after the beam was turned off. 
The difference between the two is also shown, with error bars (purely
statistical uncertainty).  There are several significant features in
the ``raw'' data from the trap-on and trap-off runs.  After the
neutron beam has been turned off, both signals approach a level of
$\sim 2$~s$^{-1}$.  This component of the background is referred to as
the ``flat'' background, since it is relatively constant over the
course of a given run.  At earlier times in the run, a time-varying
background can be seen as well.  Each of these two components of the
background arises from different sources.

The flat background in the earliest runs of the cycle was
1.4~s$^{-1}$, but rose quickly to 1.8~s$^{-1}$ by the second day, and
averaged 2~s$^{-1}$ over the entire data set.  This shift was due to
the slow buildup, from neutron activation, of radioactive nuclei with
lifetimes greater than one day.  Approximately 0.6~s$^{-1}$ of the
flat background arises from such long-lifetime activation.  The
remaining 1.4~s$^{-1}$ of the flat background probably arises from
both gamma rays penetrating the lead shielding around the cryostat and
any long-lived radioactive isotopes (such as uranium or thorium)
naturally occurring in the materials of the apparatus or shielding. 
Without further measurements, the possibility that the majority of the
present ``flat'' background comes from naturally occurring radioactive
isotopes in our apparatus cannot be ruled out.

The time-varying component of the background could include the
activation of shorter-lived isotopes, either main constituents of the
apparatus or trace impurities in or near the detector.  No single
isotope appears to dominate this component of the background, as it
did not fit well to a single exponential.  Any remnant of the
luminescence signal (discussed below) will also decrease with time,
though this signal should be eliminated by the multiphoton threshold
and requirement of coincidence between two PMTs.

Neutron absorption in the hBN (and perhaps other materials) creates
color centers which relax by emitting light\cite{Huf01}.  This
luminescence signal varies with time, temperature and magnetic field,
and is therefore the most problematic of our backgrounds.  Cuts which
were required to reduce the luminescence background also eliminate
about two-thirds of the events arising from trapped neutrons.  In
addition, the temperature dependence of the luminescence makes the
``warm'' data, in which the helium was warmed to 1.2~K, potentially
suspect (although unlikely to be so).  Thus, because of the
luminescence, data in which the helium were doped with $^{3}$He was
taken to confirm the presence of trapped neutrons.

Since the luminescence most likely results from the recombination of
electrons and holes in the boron nitride, it should emit single
photons uncorrelated in time.  Neutron decays on the other hand result
in several photons being detected in coincidence between the two PMTs,
providing an easy way to distinguish the neutron decay signal from the
luminescence background.  In practice, the rate of the luminescence
signal is more than five orders of magnitude greater than the trapping
signal.  With a single-photon counting rate of 50,000~s$^{-1}$ in each
PMT, the accidental coincidence rate (with a coincidence window of
43~ns) is 100~s$^{-1}$, three orders of magnitude larger than the
expected neutron decay signal.  Even the probability of three single
photons arriving independently within the coincidence window is larger
than the expected trapped neutron signal.  The luminescence background
can be reduced to a tolerable level (assuming that the luminescence
photons are always uncorrelated) by requiring coincidence between four
photons (corresponding to less than two false coincidences in each 1~h
run).  Recalling that a neutron decay in the center of the trapping
region emitting a 330-keV electron will produce on average a signal of
five photoelectrons, a 4 p.e. threshold is a strenuous constraint.

In practice, such a reduction in the luminescence background cannot be
obtained by setting a four-photoelectron threshold on a single
photomultiplier tube because there is not a sharp distinction between
a three- and a four-photoelectron event.  Even with a higher
threshold, a single photoelectron event can result in a much larger
``afterpulse'' in the PMT\cite{Mul52,Mor67}.  This afterpulse results
from contamination of the PMT vacuum over time (especially by helium,
which diffuses through glass) and occurs between 400~ns and 600~ns
after the initial pulse.  The size of the afterpulsing effect was
studied by illuminating a PMT with light from an LED (light-emitting
diode) at low current (and thus, producing predominantly uncorrelated
single photons).  From the pulse-height distribution, it is determined
that between 2 \% and 4 \% of the presumably single photon pulses
generated a multiple photon afterpulse.  To eliminate these
backgrounds, two separate photomultiplier tubes in coincidence are
used, with a threshold of two photoelectrons on each.

\begin{figure}[tb]
	\begin{center}
		\includegraphics{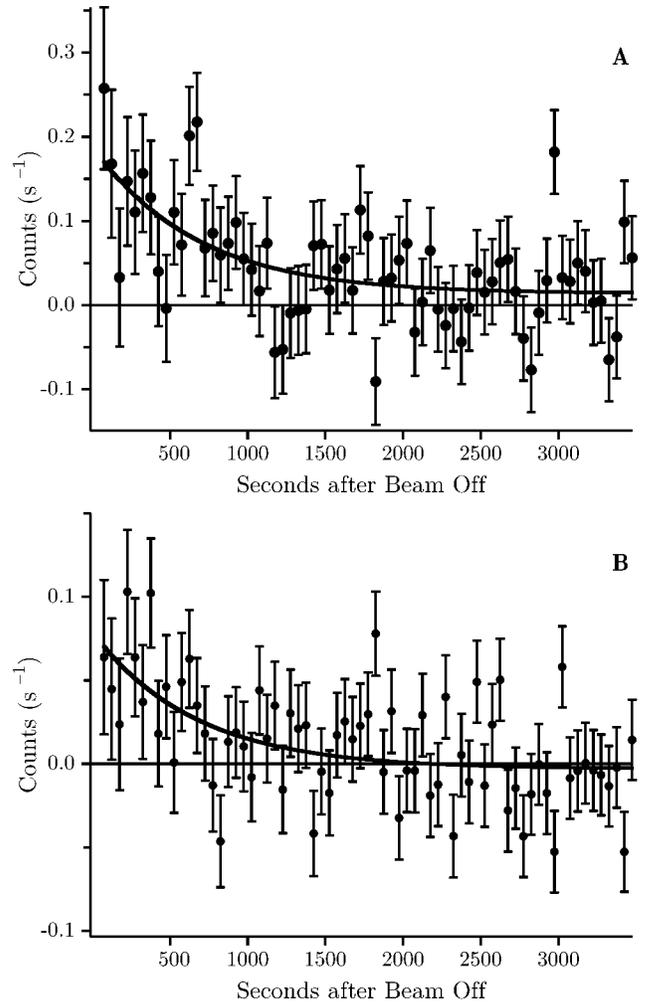}
	\end{center}
	\caption{Plot of the difference between the averaged
	``trap-on'' and the averaged ``trap-off'' data for reactor
	cycle one (\textbf{A}) and two (\textbf{B}).  The solid lines
	indicate the result of the fits described in the text.}
	\label{fig:differencedata}
\end{figure}

Fig.~\ref{fig:differencedata} shows the pooled background subtracted
data; i.e.\ the difference between the pooled trap-on data and the pooled
trap-off data, divided by the number of runs which were pooled
together.  The following models are simultaneously fit to the data:
\begin{eqnarray}
    \label{eq:datafit}
	W_{1} &=& A_{1} e^{-\frac{t}{\tau}} + C_{1} \\
	W_{2} &=& A_{2} e^{-\frac{t}{\tau}} + C_{2}, \nonumber
\end{eqnarray}
where $W$ is the experimental counting rate and the subscripts
correspond to the first and second reactor cycle.  The $A$'s
correspond to the amplitude of the neutron decay signal, $\tau$
corresponds to the lifetime of the neutrons in the trap (assumed to be
the same in each reactor cycle) and the $C$'s correspond to any
residual offset between trap-on and trap-off runs, which could arise
from small changes in the ``flat'' background rate over the course of
many runs.

The data from each cycle are binned into seventy 50~s bins, with the
first starting 50~s after the beam is turned off.  The five
parameters, $A_{1}, A_{2}, C_{1}, C_{2},$ and $\tau$ are varied using
a routine adapted from the grid fit algorithm described in ref.\
\cite{Bev92} to minimize the chi-squared ($\chi^{2}$) of the fit.  The
values of the fit parameters are those which yield the minimum
$\chi^{2}$, and the 68~\% confidence interval for each parameter is
given by the range over which the $\chi^{2}$ (varying the other four
parameters) increases by one relative to the minimum (varying all five
parameters).

The fit gives a minimum $\chi^{2}$ of 150.67 with 135 degrees of
freedom at\cite{NatureNote}:
\begin{eqnarray}
	A_{1}&=&  0.175 \mathrm{~s}^{-1} \pm{0.045} \mathrm{~s}^{-1}\nonumber\\
        A_{2}&=&  0.082 \mathrm{~s}^{-1} \pm{0.021} \mathrm{~s}^{-1}\nonumber\\
	C_{1}&=&  0.014 \mathrm{~s}^{-1} \pm{0.011} \mathrm{~s}^{-1}\label{fitvals}\\
	C_{2}&=& -0.003 \mathrm{~s}^{-1} \pm{0.005} \mathrm{~s}^{-1}\nonumber\\
	\tau &=&  660   \mathrm{~s}~^{+290}_{-170} \mathrm{~s}.\nonumber
\end{eqnarray}
Note that $A_{1}$ and $A_{2}$ are significantly above zero, 
indicating trapped UCN are present in the trap.

The data shown in Fig.~\ref{fig:differencedata} are evidence of an
increase in the counting rate in runs in which the magnetic trap is
turned on.  However, from this evidence alone we cannot definitely
conclude that the increase represents an additional source of counts
from trapped UCN as opposed to, for example, small, magnet-dependent
changes in the rate of neutron-induced backgrounds.  It has been
assumed that the trap-on data contain both the neutron decay signal
and background events, while the trap-off data contain no decay
events, but the same backgrounds.  If this assumption is true, then
trap-on and trap-off data taken under conditions chosen to eliminate 
the trapped UCN without changing the backgrounds should be consistent 
with each other.

To keep the background rate the same in any different experimental
configuration, the detector (including the scintillation properties of
the helium bath are concerned), must remain unchanged.  Similarly, the
neutron beam, which causes all the time-varying backgrounds, must
remain unchanged as well.  Two changes to the experimental
configuration which meet these conditions and remove the trapped
neutrons are warming the helium target to a temperature of 1.2~K and
doping a small amount of $^{3}$He ($x_{3}\approx 10^{-7}$) into the
isotopically purified $^{4}$He.

In ref.\ \cite{Gol83}, the upscattering rate for UCN stored in
superfluid helium was measured to be $2.5\times 10^{-2}$~s$^{-1}$ at
1.16~K, giving a UCN storage lifetime of 40~s at that temperature. 
During the loading phase, the continuous upscattering of UCN reduced
the expected peak number of trapped UCN in our experiment from $P
\tau_{\beta}$ to $P \tau_{up}$, a factor of
$\tau_{up}/\tau_{\beta}\approx 0.05$.  With $P=0.73$ (for the
$E_{T}=0.8$~mK runs), this gives a maximum of 29 trapped UCN when the
beam is turned off, or an initial beta-decay rate of 0.03~s$^{-1}$,
decreasing with a lifetime of 40~s, to under 0.01~s$^{-1}$ before the
first analyzed data, 50~s after the beam is turned off.  This is
compared to an expected rate of 0.16~s$^{-1}$ at that time in the
trapping data.  Since the luminescence background varies with
temperature, as well as with magnetic field, the warm runs could
conceivably produce a false negative.  The $^{3}$He technique does not
have this feature, making it a more conclusive negative than the
``warm'' data.

For the $^{3}$He runs, 10~Pa-L of $^{3}$He was introduced into the
$8\times10^{7}$~Pa-L of $^{4}$He filling the experimental cell,
yielding a concentration $x_{3} \equiv
n(^{3}\mbox{He})/n(^{4}\mbox{He}) \approx 10^{-7}$.  This
concentration gives a UCN lifetime of 0.4~s and an absorption length
of 400~m for 0.395~nm neutrons, the lowest wavelength neutrons passing
through the beryllium filter.  This attenuation is a factor of two
shorter at 0.8~nm, on the tail of the spectrum.  Over the 1~m length
of helium in the target, the $^{3}$He should
absorb less than 1~\% of the beam.  Thus, any background arising from
the cold neutrons in the beam, rather than from UCN remaining after
the beam is turned off, should be changed by less than one percent.

\begin{figure}[tb]
	\begin{center}
		\includegraphics{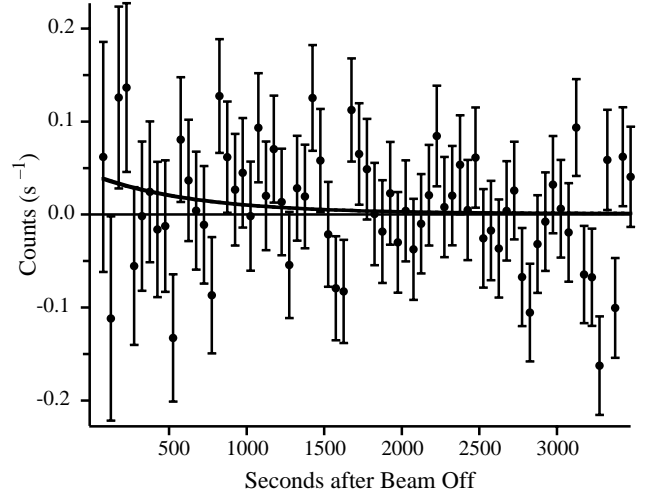}
	\end{center}
	\caption{Plots of the difference between the average pooled
	magnetic ``trap-on'' and ``trap-off'' in the 1.2~K ``warm''
	data taken in reactor cycle 1.}
	\label{fig:warmdata}
\end{figure}

The data from the warm runs are shown in Fig.~\ref{fig:warmdata}. 
There are no warm data from the second reactor cycle, so the data are
fit only to $A_{1}$ and $C_{1}$, with $\tau=660$~s held fixed.  This
two parameter fit gives a result of:
\begin{eqnarray}
	A_{1} &=& 0.042 \mathrm{~s}^{-1} \pm {0.043}\mathrm{~s}^{-1}\nonumber \\
	C_{1} &=& 0.001 \mathrm{~s}^{-1} \pm {0.009}\mathrm{~s}^{-1}\nonumber,
\end{eqnarray}
with a $\chi^{2}$ of 76.1 for 68 degrees of freedom.  This is 
consistent with the prediction of $A_{1} \lessapprox 0.01$~s$^{-1}$.

\begin{figure}[tb]
	\begin{center}
		\includegraphics{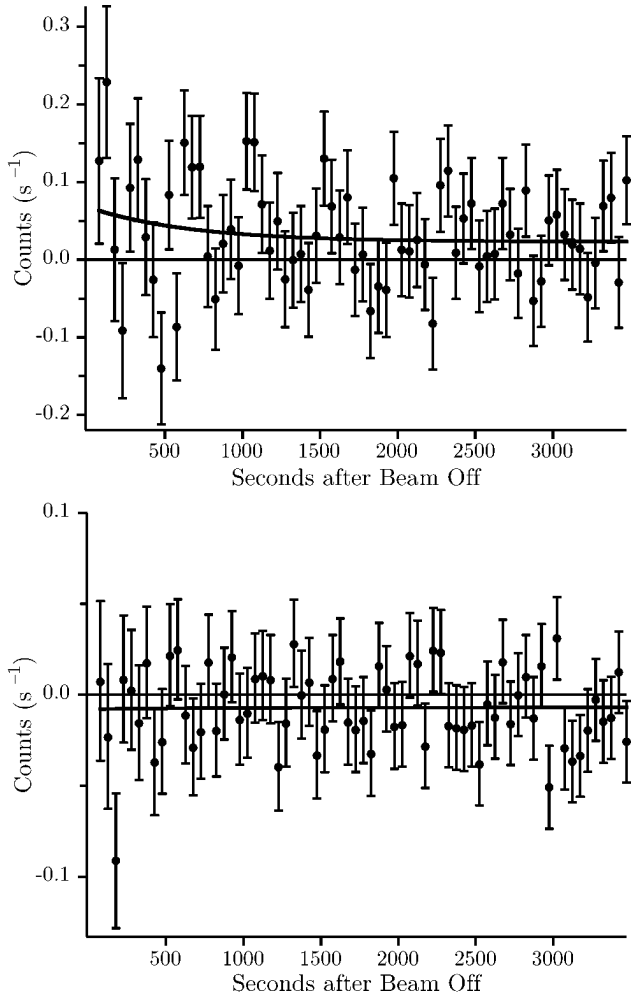}
	\end{center}
	\caption{Plots of the difference between the average pooled
	magnetic ``trap-on'' and ``trap-off'' in the $^{3}$He data for
	reactor cycle one (\textbf{A}) and two (\textbf{B}).}
	\label{fig:heliumdata}
\end{figure}

The data from the $^{3}$He runs are shown in Fig.~\ref{fig:heliumdata}. 
The data are fit to eq.~\ref{eq:datafit} with $\tau = 660$~seconds. 
This four parameter fit gives the following results, with a minimum
$\chi^{2}$ of 119.3 for 136 degrees of freedom:
\begin{eqnarray}
	A_{1}&=&  0.045 \mathrm{~s}^{-1} \pm{0.041} \mathrm{~s}^{-1}\nonumber\\
        A_{2}&=& -0.001 \mathrm{~s}^{-1} \pm{0.017} \mathrm{~s}^{-1}\nonumber\\
	C_{1}&=&  0.023 \mathrm{~s}^{-1} \pm{0.009} \mathrm{~s}^{-1}\nonumber\\
	C_{2}&=& -0.007 \mathrm{~s}^{-1} \pm{0.0035}\mathrm{~s}^{-1}\nonumber.
\end{eqnarray}
The amplitudes $A_{1}$ and $A_{2}$ are consistent with zero trapped UCN
in the trapping region and inconsistent with the signal observed in
the trapping data.  The remaining constant offsets arise from the slow
change in the ``flat'' background rate from run to run.
A discussion of these results is given below.

\section{Discussion}
\label{sec:disc}
For an infinite loading time, the expected initial number of trapped neutrons
is given by $P \tau_{n}$, where $P$ is the production rate of trapped
UCN and $\tau_{n}$ is the storage time.  For a finite loading time,
the initial number of trapped UCN, $N$, is given by:
\begin{equation}
	N = P \tau_{n} (1-\exp{(-t_{\mathrm{load}}/\tau_{n})}).
\end{equation}
The production rate of trapped UCN, $P$, for the 0.75~mK runs, is
theoretically predicted to be $0.73~\mathrm{s}^{-1}\pm
0.24$~s$^{-1}$\cite{Bro00}.  If beta decay is the only significant
loss, then $\tau_{n}=\tau_{\beta}=886.7$~s\cite{PDG}.  Based on the 
predicted production rate, a loading
time of $t_{\mathrm{load}}=1350$~s gives $N_{1} = 500 \pm 170$.  In
the 0.5~mK runs the production rate, $P$, and hence $N_{0}$, is scaled
by a factor of (120~A/180~A)$^{3/2} = 0.544$, giving $N_{2} = 270 \pm 90$.

The initial count rate in our detector is given by $W=(\epsilon
N)/\tau_{\beta}$, where $\epsilon=31~\%$ is the detector efficiency. 
The number of trapped neutrons observed in each data set is obtained
by taking $\hat{N}_{i}=\tau_{\beta} A_{i} / \epsilon$.  For the first
data set, $A_{1} = 0.175 \pm 0.045$ gives $\hat{N}_{1} = 500 \pm 155$. 
For the second data set, $A_{2} = 0.082 \pm 0.021$ gives $\hat{N}_{2}
= 230 \pm 70$.  Thus, the fit to each data set is consistent with the
expected number of trapped UCN\@.  In addition, when the amplitudes of
the trapping and non-trapping (``warm'' and ``$^{3}$He) runs are
combined, the trapping runs have an amplitude which is three standard
deviations larger than the amplitude from the non-trapping runs.  This
constitutes strong evidence that we have magnetically trapped
neutrons.

\section{Conclusions and Future Directions}
\label{sect:conclusion}

Having demonstrated magnetic trapping of UCN, we now consider the
prospects for making a measurement of the neutron lifetime.  First,
the error bars on the measurement quoted in eq.~\ref{fitvals} may seem
discouraging, compared to the weighted world average value of $886.7
\mathrm{s} \pm 1.9$~s\cite{PDG}.  However, the uncertainty in the
present data is due entirely to statistical fluctuations in the large
backgrounds being subtracted and not from the number of trapped UCNs.

There is every reason to believe that a large reduction in at least
the time-varying background is possible, since these backgrounds are
created by neutrons of all wavelengths, while the trapped UCN are
produced by neutrons in a narrow band around 0.89~nm.  Thus, a 0.89~nm
nearly monochromatic beam should reduce these backgrounds by almost
two orders of magnitude relative to the neutron signal.

The flat background can be reduced by increasing the detection
efficiency, allowing higher rejection thresholds to be set. 
Presently, the efficiency is 31~\%, mainly limited by the transport of
the blue light down the tube surrounding the trapping region.  An
increase in the wall thickness of the tube will not only allow
efficiencies $> 90$~\%, but will allow the thresholds to be raised,
thereby lowering the backgrounds.  The detection efficiency may also
be raised by using an optically transparent beamstop rather than the
opaque B$_{4}$C beamstop.

Lastly, the number of trapped UCN could be increased by building a
magnetic trap either physically larger than that described here,
``deeper'' (in terms of magnetic trap depth), or both.  The number of
trapped UCN is approximately proportional to the volume of the trap
($l r_{T}^{2}$) and to $k_{T}E_{T} \propto B_{T}^{\frac{3}{2}}$.  In
fact, collimation constraints in setups of our type make the UCN
population depend slightly more strongly on $r_{T}$ and less so on
$l$.  The present magnet has a bore of 5.08~cm and the diameter of the
trap is 3~cm.  In a larger-bore magnet, this ratio will substantially
improve, giving a trap depth that is a larger fraction of the field at
the bore of the magnet.  An improvement of more than two orders of
magnitude in the number of trapped UCN, giving a potential measurement
at the 1~s level, is likely in the next few years.

Ultimately, this method of measuring the neutron lifetime is limited
by losses of neutrons from the trap, which would give the appearance
of a lifetime shorter than the actual beta-decay lifetime.  Ways in
which neutrons could be lost from the trap, other than beta-decay,
include: scattering by excitations in the liquid helium, absorption by
$^{3}$He, Majorana spin-flip transitions and the marginal trapping of
neutrons.  Scattering of trapped neutrons by excitations in the liquid
helium can be reduced to below $10^{-5}$ of the beta-decay rate by
cooling the helium to below 150~mK \cite{Gol79,Gol83}.  Absorption by
$^{3}$He is reduced by using isotopically purified $^{4}$He.  At the
level to which the supplied $^{4}$He has been
verified\cite{Hen87,Hen81}, a limit of $10^{-2}$ of the beta-decay
rate can be set.  The actual purity is probably better and will be
measured using accelerator mass spectrometry.  Majorana spin-flip
transitions occur near zero-field regions and can be suppressed in an
Ioffe trap by the application of a bias field.  Marginally trapped
neutrons can be eliminated by ramping the magnetic field down to 30\%
of its maximum value and then back up, resulting in a population of
UCN all with energies less than the trap depth and thus all truly
trapped.  

The beta-decay lifetime of a neutron in a bath of liquid helium could
in principle differ from that of a neutron in vacuum due to either a
change in the available phase space for the decay products or a change
in the matrix element that governs the decay.  Small changes in the
phase space factor arise from an increased or decreased amount of
energy available to the decay products.  For example, the minimum
energy of the electron in helium is 1.3~eV\cite{Som64}.  The net
effect of this energy difference is to increase the neutron lifetime
by at most $6.5\times 10^{-6}$.  We estimate that all of the
phase-space effects are less than $10^{-5}$.  Also, the matrix element
for the decay could be influenced by the presence of the helium
nuclei.  The axial-vector coupling constant $g_{a}$ is known to vary
by 26\% compared to that of muon decay\cite{Tow98}.  We can estimate
the effect of such a change by considering the range of the strong force
($\sim 10^{-15}$~m) and the density of helium nuclei in the neutron's
environment.  The effect of the helium nuclei in the environment of
the neutron should be $\sim 10^{-16}$.  There is no reason to expect
any of these effects, or the loss mechanisms mentioned above, will
prevent a measurement of the neutron lifetime at the $10^{-5}$ level,
given sufficient statistics.

In addition to the immediate use of trapped UCN in a measurement of
the neutron lifetime, the techniques developed for this experiment are
applicable to a wide range of future experiments.  The detection of
EUV scintillations in liquid helium (or neon) using wavelength
shifters has been proposed for use in a solar neutrino
experiment\cite{McK00b}.  Also, both the production and storage of UCN
in a helium bath and the detection techniques described herein are
relevant to a proposed experiment to measure the electric dipole
moment (EDM) of the neutron using polarized UCN in a bath of liquid
helium\cite{Gol94}.

The apparatus described above could also be readily adapted for use as
a relatively strong source of polarized UCN\@.  This source of
perfectly polarized UCN could be useful in the determination of
correlation coefficients in neutron decay which can be combined with
$\tau_n$ to test the Standard Model.  A UCN beam could
also be used as a probe for the investigation of condensed matter,
because the UCN's wavelength is comparable to interesting correlation
lengths in crystalline, polymeric and biological materials.

\section*{Acknowledgments}
We thank J.\,M.~Rowe, D.\,M.~Gilliam, G.\,L.~Jones, J.\,S.~Nico,
N.~Clarkson, G.\,L.~Yang, G.\,P.~Lamaze, C.~Chin, C.~Davis, D.~Barkin,
A.~Black, V.~Dinu, J.~Higbie, H.~Park, R.~Ramakrishnan, I.~Siddiqi,
B.~Beise and G.~Brandenburg for their help with this project.  We
thank P.~McClintock, D.~Meredith and P.~Hendry for supplying the
isotopically pure helium.  This work is supported in part by the
National Science Foundation under grant No.  PHY-9424278.  We
acknowledge the support of the NIST, US Department of Commerce, in
providing the neutron facilities used in this work.  The NIST authors
acknowledge the support of the US DOE.


\begin{thebibliography}{10}
\bibitem{NatureNote}
The fit values here, and for the $^{3}$He runs differ
from those cited in reference \protect\cite{Huf00a}, because several sets of
data, excluded in reference \protect\cite{Huf00a} are included in the fits
here.  The excluded data are from the first six hours of data (four
runs) after any long period in which the apparatus is not irradiated. 
The flat background is changing rapidly between these runs, giving a
non-zero $C_{1}$ and $C_{2}$ in the fits shown here.
\end{thebibliography}
\end{document}